\documentclass[]{aa}  
\usepackage{graphicx}
\usepackage{txfonts}
\usepackage{natbib}
\begin{document}
\title{On planetary mass determination in the case of super-Earths orbiting active stars. 
The case of the CoRoT-7 system.}
\author{S. Ferraz-Mello
     \inst{1}
     \and
      M. Tadeu dos Santos\inst{1}
     \and
      C. Beaug\'e\inst{2}
     \and
      T.A. Michtchenko\inst{1}    
     \and
      A. Rodr\'{\i}guez \inst{1}}    
\institute{Instituto de Astronomia, Geof\'isica e Ci\^encias Atmosf\'ericas (IAG) -- Universidade de S\~ao Paulo\\
              Rua do Mat\~ao, 1226 Cep: 05508--090  - S\~ao Paulo-Brasil\\
              \email{sylvio@astro.iag.usp.br} \\
         \and
             Observat\'orio Astron\'omico de C\'ordoba, Universidad Nacional de C\'ordoba\\
             Argentina}
%
%
\abstract
    {Due to the star activity, the masses of CoRoT-7b and  CoRoT 7c are uncertain. Investigators of the CoRoT team have proposed several solutions, all but one of them larger than the initial determinations of  $4.8 \pm 0.8 \,M_{\rm Earth}$ for CoRoT-7b and $8.4 \pm 0.9 \,M_{\rm Earth}$ for CoRoT 7c. }
     {This investigation uses the excellent HARPS radial velocity measurements of CoRoT-7 to re-determine the planet masses and to explore techniques able to determine mass and elements of planets discovered around active stars when the relative variation of the radial velocity due to the star activity cannot be considered as just noise and can exceed the variation due to the planets. }     
     {The main technique used here is a self-consistent version of the high-pass filter used by Queloz et al. (2009) in the first mass determination of CoRoT-7b and CoRoT-7c. The results are compared to those given by two alternative techniques: (1) The approach proposed by Hatzes et al. (2010) using only those nights in which 2 or 3 observations were done; (2)  A pure Fourier analysis. In all cases, the eccentricities are taken equal to zero as indicated by the study of the tidal evolution of the system; the periods are also kept fixed at the values given by Queloz et al. Only the observations done in the time interval BJD 2,454,847  -- 873 are used because they include many nights with multiple observations; otherwise it is not possible to separate the effects of the rotation fourth harmonic (5.91 d=$P_{\rm rot}/4$) from the alias of the orbital period of CoRoT-7b (0.853585 d). }                              {The results of the various approaches are combined to give for the planet masses the values $8.0 \pm 1.2 \, M_{\rm Earth}$ for CoRoT-7b and $13.6 \pm 1.4 \, M_{\rm Earth}$ for CoRoT 7c. An estimation of the variation of the radial velocity of the star due to its activity is also given. }
    {The results obtained with 3 different approaches agree to give masses larger than those in previous determinations. From the existing internal structure models they indicate that CoRoT-7b is a much denser super-Earth. The bulk density is $11 \pm 3.5 \,{\rm g.cm}^{-3}$. CoRoT-7b may be rocky with a large iron core.}


\titlerunning{Planetary mass determination of super-Earths orbiting active stars}

\maketitle
\section{Introduction}
\begin{figure}
\resizebox{\hsize}{!}{\includegraphics{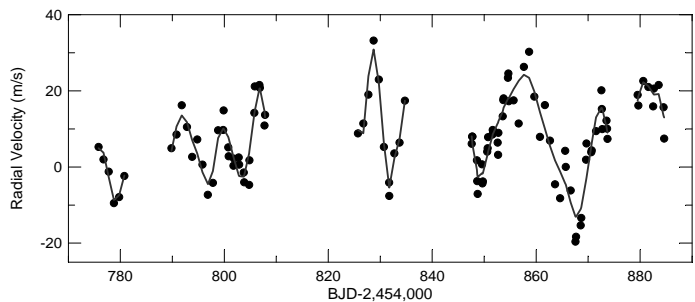}}
\caption{Relative radial velocity measurement (dots) and the contribution of the activity as estimated with a filtering using the first three harmonics of the rotational period (solid line).(Adapted from Fig.8 of QBM)}
\label{fig:fig1}
\end{figure}

CoRoT-7b was the first super-Earth for which mass and radius have been determined.
CoRoT-7b and the recently discovered GJ 1214b (Charbonneau et al.  \citep{charbonneau}) and Kepler-10b (Batalha et al.  \citep{batalha}) are paradigms
for the study of the physics of what exo-Earths, super-Earths and/or mini-Neptunes can be. 
They set the only real constraints available to models of the formation
and evolution of hot telluric planets. For this reason, it is very important to have good
radius and mass determinations. The radius of CoRoT-7b, determined from the transits
observed by CoRoT\footnote{The CoRoT space mission, launched on December 27th 2006, has been developed and is operated by CNES, with the partnership of Austria, Belgium, Brazil, ESA, Germany and Spain.}, is 10,100 $\pm$ 600 km (Bruntt et al.  \citep{bruntt}), a value that may be
improved, but whose magnitude is nevertheless definitively established. The mass,
 {determined from the radial velocity measurements} ($4.8\pm 0.8$ Earth masses cf Queloz et al.  \citep{queloz}; hereafter QBM), however, has
not the same accuracy. This is due to the fact that the $0.91 \pm 0.03 M_{\odot}$ (see Lanza et al.  \citep{lanza}), 1-2 Gyr-old G9V star CoRoT-7
(=TYCHO 4799-1733-1) is too active. The variation in the measured radial velocity comes
mainly from the activity of the star whose spots determine the value of the radial velocity
integrated over its disk. This activity follows roughly the rotation period of the star (23.64
days), but in a very irregular way: the 150-day photometric observations done by CoRoT
show the variation displaying full span in some times, but almost disappearing in others
(see L\'eger et al. \citep{leger}). Because of this activity, CoRoT-7b is also the paradigm of the kind
of problems that may be often found when planets with Earth-like masses are discovered. One
of the aims of the investigation reported in this paper is to use CoRoT-7, for which an
excellent set with 109 HARPS radial velocity measurements exists, to explore techniques
that may be used when dealing with low-mass planets. We may wish that future discoveries
are done around more quiet stars, but Earth-like planets are too important and we cannot
discard any of them because of the central star activity. We hope that CoRoT, KEPLER and
ground based instruments will discover new Earths, super-Earths and mini-Neptunes which, { as Kepler-10b, may be in orbit around quiet stars. 
However,} as CoRoT-7b, they may be found around 
non-quiet stars and the improvement of the techniques of
mass and orbit determination used to study such cases is important. This need is motivating
a great deal of investigations and the number of papers dealing with the mass of CoRoT-7b
(QBM; Boisse et al.  \citep{boisse}; Hatzes et al.  \citep{hatzes}; Pont et al. \cite{pont} ) is increasing. In addition to
this interest, we have to consider that HARPS, currently the only instrument able to make
measurements of CoRoT-7 with the required precision, is being on demand by a great deal
of other targets and we cannot foresee when a new series of measurements with the same
quality of the existing one will become possible.\\
The nature and magnitude of the problem of the mass and orbit determination of the
CoRoT-7 planets can be assessed from fig.~\ref{fig:fig1} (adapted from fig. 8 of QBM). It shows the
5 sets of measurements of the radial velocity of CoRoT-7 obtained with HARPS (dots) and
the estimated part of the radial velocity due to the star activity. The activity shown in fig.~\ref{fig:fig1}
(solid line) was obtained in QBM by means of a filtering
designed to eliminate periodic disturbances corresponding to the first three harmonics\footnote{We follow the common usage in Physics where $N^{th}$ harmonic means an oscillation whose frequency is $N$ times the fundamental frequency. Thus the fundamental frequency $2\pi/P_{\rm rot}$ is the first harmonic, the second harmonic is the component whose period is $P_{\rm rot}/2$, and so on.} 
of the rotational period. These sets present patterns very different one from another showing
that the construction of one single model for the activity in the whole interval is not
possible. It also discourages the extensive use of Fourier tools (periodograms) over the
whole set.
We have to search for techniques able to separate the low-frequency rotation signals
(periods 23.64 days and its main harmonics), from the higher frequency signal coming from
the two planets (periods 0.853585 and 3.698 days cf. QBM). The difference between the
periods of the two components (rotation and planets) indicates that, in this case, we may
filter the data from its low-frequency parts and consider separately the high-frequency
information. In the case of one continuous signal, or at least of a long discrete evenly
spaced time series, the problem that we have to solve is classical and well known: We
should construct and use a high-pass filter. However, the simple recipes to construct a high-pass
filter in the frequency domain cannot be used for a series with a limited amount of data and,
worse, unevenly distributed. Because of the high correlation between the absolute values of
the Fourier transform (or spectral power) at different frequencies, these filters
must necessarily be constructed in the time domain.\\
One high-pass filter in the time domain was used in QBM in the first mass determination of
CoRoT-7b and-7c. However, as discussed in Section 2 of this paper, the filter then used
affected the high-frequencies, the importance of which was downsized without apparent
reason, and set the cut-off below the frequency of the main alias of the period of CoRoT-
7b, thus affecting also the amplitudes corresponding to this planet (cutting out the alias, the
filter also considerably downsized the signal corresponding to the actual frequency of the
planet).\\
This paper starts with an analysis of the mass determination published in QBM's CoRoT-7c
discovery paper (section 2) and then proposes a self-consistent açgorithm founded on the filtering
technique used
there (section 3), which is applied to the selected set of 52 observations made in 27
consecutive nights, between BJD 2,454,847 and BJD 2,454,873 (section 4) to obtain the
masses of the planets. 
 {The restriction of the analysis to this set of dates is due to the fact that, in 10 nights of this set, 3 observations were done, spanning about 4 hours between the first and last observation in the night. This is not enough to completely destroy the aliasing due to the almost uniform spacing between observations done in consecutive nights, but it allows us to distinguish between two solutions with forced periods equal to the transit period and its alias (see Hatzes et al.  \citep{hatzes} figure 7). It is worth stressing the fact that in a series made of observations taken always near the same hour in the night, no mathematical tool exists able to distinguish between one frequency and its aliases.} 
In section 4, the dependence of the results on the filter parameters is also discussed. 
In sections 5 to 7, we present the resulting estimate of the
star activity and analyze the residuals obtained by subtracting the activity from the
observed radial velocities. Alternative techniques are discussed in sections 8 and 9. 
textbf {
The approach discussed in section 8 follows a sugestion by Hatzes et al.  \citep{hatzes} and uses only the observations from nights where multiple observations were done. 
This is of particular importance because it allows an analysis independent of any explicit hypotheses on the behavior of the star activity. These observations are analysed here with the help of a biased Monte-Carlo technique allowing confidence intervals to be obtained.
The approach discussed in section 9 is a classical multi-period Fourier analysis. 
It differs from other approaches using a Fourier decomposition by the fact that, here, no a priori periods are used. The periods of the solution found are those allowing us to get the best fit of the observations to a multi-periodic function.}
In section 10, we present some
simulation results taking into account tidal interactions showing the circularization of the
orbits and thus justifying the adoption of zero eccentricities for both planets. At last, we
proceed with the discussion of the results and the conclusions.
\section{Analysis of the first mass determination}
\begin{figure}
\resizebox{\hsize}{!}{\includegraphics{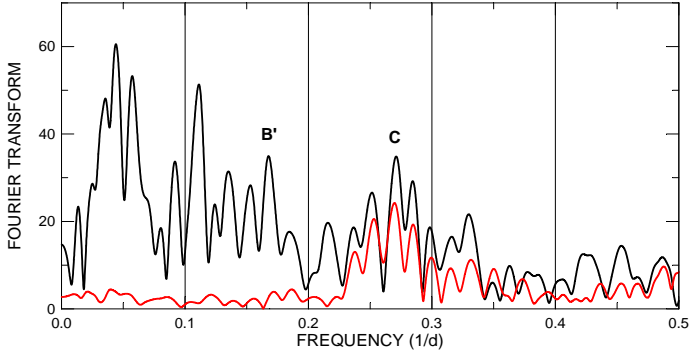}}
\caption{Fourier transform (DCDFT) of the measured radial velocities (black) and
of the values obtained in QBM after subtracting the estimated activity (red). Both
transforms are presented using the same units. C and B' indicate the orbital frequency
of CoRoT-7c and the 5.925-day alias of the CoRoT-7b frequency, respectively}
\label{fig:fig2}
\end{figure}

The analysis of the mass determination published in the CoRoT-7c discovery paper (QBM)
is the first step in this study and the main point to be considered concerns the filtering
properties of the procedure used there. Is it equivalent to a high-pass filter? In order to
know that, we compute the Fourier transforms \footnote{Spectra obtained using date-compensated discrete Fourier transforms (DCDFT; cf Ferraz-Mello  \citep{sylvio}). DCDFT differs from usual Lomb periodograms because they also consider the constant component, whose neglect
may affect the height of the peaks (see discussion on floating-mean periodograms in Cumming et al. \cite{cumming} )} of the given data and of the residuals obtained in QBM after subtracting the activity, respectively, and compare them one to
another. The result presented in fig.~\ref{fig:fig2} shows that, indeed, the used procedure 
completely filtered the low-frequencies (the transform of the filtered series is close to zero for all frequencies below 0.22 $d^{-1}$. However, it also affected the high-frequency components.
The strong alias of the period of CoRoT-7b, at 5.925d (indicated with B' in fig.~\ref{fig:fig2}),
disappeared. The signal of CoRoT-7c was less affected by the filter but it was downsized
without apparent reason, since no low-period terms were included in the filter. 
The main reason for this result is the fact that the filtering was done on the raw data,
without taking into account the part of the signal due to the planets. We may guess that the
fact that these periods are not at all commensurable with the Coherence Time (window)
used in the filter (20 days) may have played a role in the deep sculpting done by the
filter at the planet frequencies and their aliases { (the average of a periodic function over a time interval not commensurable with its period is different of zero.)}.\\
A second factor may have been the fact that fittings are exact when the number of
dates is smaller (as well known). Thus, near the borders of the interval, the estimated
activity will be closer to the given data than in the middle of the period. The end effect seems to be 
responsible for the fish-like appearance of the filtered velocities in the largest set (narrow in the
extremities and wide in the middle; See the grey line and dots in fig.~\ref{fig:fig8}). One could think
that such appearance might result from a particular beat of the orbital frequency of CoRoT
7c and the nearby alias of the orbital frequency of CoRoT-7b, but an a posteriori plot of
synthetic velocities shows that this is not so. Because of the actual phases and periods, the
destructive interference of the two sine curves (at the actual discrete observation times)
does not happen in the borders of the interval, but near the middle of it.\\
It is worth adding that one unconstrained 3-sinewave analysis of the filtered radial velocities in QBM, in the period
BJD 2,454,847 –- 873, using a genetic algorithm completed with a downhill simplex gave as
more important periods present in the filtered data, 3.495d and 3.963d. The difference in
frequency of these two periods is $0.0338 d^{-1}$, which is the inverse of 29.6 d (very close to
the timespan of the observations used in the analysis), clearly showing the interference of
the timespan of the observations in the considered subset.
\section{A self-consistent high-pass algorithm}
\begin{figure}
\resizebox{\hsize}{!}{\includegraphics{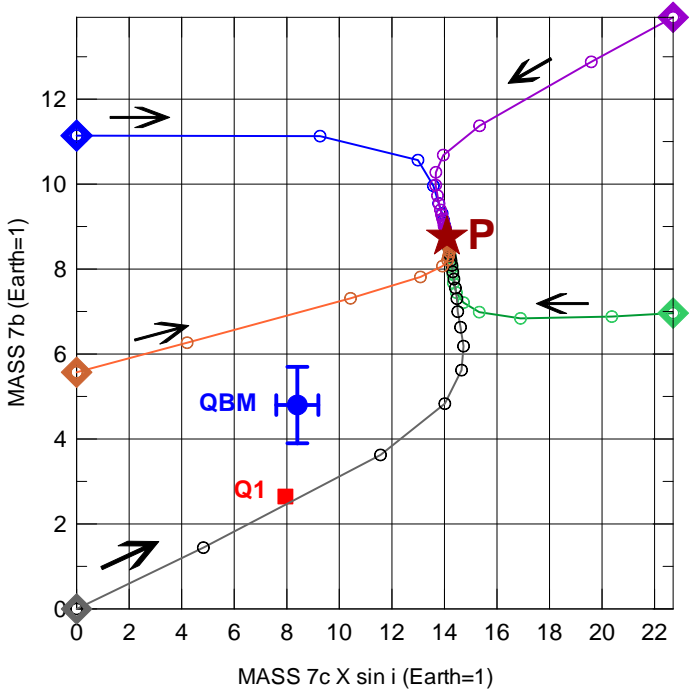}}
\caption{Evolution of the masses in five chains of results iteratively obtained from
{ five different sets of initial conditions (diamonds) using the 4-harmonic high-pass filter and observations done in the period BJD 2,454,847 -- 873. All of them converge to the same point $P$ (star).}
One of the chains started at (0,0). The figure also
shows the result of the first iteration in QBM (point Q1) and the mass values adopted in that paper. }
\label{fig:fig3}
\end{figure}

\begin{figure}
\resizebox{\hsize}{!}{\includegraphics{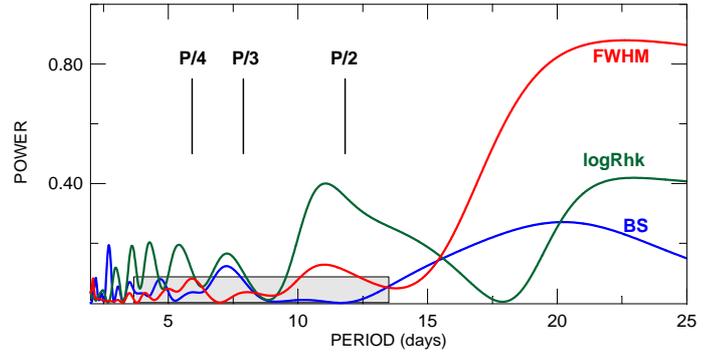}}
\caption{{Power spectra of the spectrographic
parameters associated with the star activity: bisector span, FWHM and $\log R_{hk}$, in the
period BJD 2,454,847 -- 873. The exact location of the main rotation harmonics is indicated.
The gray rectangle is below the minimum significance level. 
Peaks inside this rectangle can be produced by random data.}}
\label{fig:fig4}
\end{figure}

The high-pass filter used in QBM may be shortly described as follows. First, we define one
time window (the Coherence Time), fixed as being a guess on the number of days in which
it may be reasonable to fit the given harmonic function to the activity. Then, we construct N time
windows of the chosen size, each centered on one of the N dates of the observations and
including all observations inside the window. In every window, the data are Fourier
analyzed and represented by the first terms of a harmonic series whose fundamental period
is the star rotation period (23.64 d) {plus a constant}. The activity at a given date is estimated as the (weighted)
average of the values given to it by the harmonic representations of the signal in all subsets
including the given date. For the sake of clearness, let us add the following information: (i)
The actual window size is fixed in such a way that the ends of the window do not
separate observations done in the same night (if necessary, the actual window is taken
slightly larger than the nominal Coherence Time); (ii) Near the borders, the subsets are
incompletely filled as the windows extend to beyond the considered interval covering some
nights where no observations were done.\\
In this paper, we propose an improvement of this procedure. The main change is that,
now, a predicted signal corresponding to the two planets is subtracted from the radial
velocities before the fitting of the harmonic function. Next, the activity is estimated as
described above and subtracted from the observed radial velocities; the resulting residuals
are used to determine the masses. The new masses may then be used as first guesses in a
new run of the algorithm and leads to an improved prediction of the signal corresponding
to the two planets, and so on. The procedure is iterated as many times as necessary up to
reach a satisfactory convergence. Formally, we may say that the whole procedure defines a
map $x_{n+1} = F(x_n)$, (here, x represents the masses) which is iterated up to get $x_{n+1}=x_n$. One
problem appearing in the actual application of this scheme is a possible slow convergence of
this map. For this reason, to obtain the results discussed in this paper, we have rather used an
alternative accelerated map: $x_{n+1}=F(x_{n}+\lambda(x_n-x_{n-1}))$ with $\lambda=1$.\\
In a self-consistent solution, the values of the masses used, to subtract the planets, is equal to that 
obtained from the flitered series. Figure~\ref{fig:fig3}  shows the evolution of
the masses obtained in five chains of determinations using the same high-pass filter (with 4
harmonics and the same Coherence Time). The longest one was obtained iterating the
process from a first approximation in which the actual observed radial velocities are used without taking into account the planets
(i.e. starting with the two masses equal to zero). The others started from arbitary sets of values corresponding to higher masses, represented by diamonds in fig. \ref{fig:fig3}. 
 One may note the very slow convergence of the iterative procedure. 
{ One may also note that the algorithm converges quickly to a line that corresponds to a linear relation between the two masses in which one of them ($m_{\rm 7c}$) is almost constant. To have different convergence ratios along two orthogonal directions is a common feature in maps; in this case it seems to be due to the slow separation between the planet CoRoT 7b and the fourth rotation harmonic. However, notwithstanding the slow convergence, the five runs converged to the same point $P$, represented by a star in Fig. \ref{fig:fig3}} 
{The reliability of the map was checked using some synthetic sets of data constructed using the first, second and fourth rotation harmonics with the amplitudes indicated in some Fourier analyses (see section 9), two sine curves of half-amplitude 6 m.s$^{-1}$, corresponding to two planets, and a Gaussian noise. The results are as follows: (1) When the noise is not included, the map reproduces the two planets exactly as given; (2) When noise in the range 1.7--2.5 m.s$^{-1}$ is added, the results for CoRoT-7b fall around the given value within 0.3 m.s$^{-1}$. However, the results for CoRoT-7c fall systematically $\sim$1 m.s$^{-1}$ below the given value. We have taken these results into account in estimating the error bars of the final results.} \\
 {The analysis of the spectrographic parameters related to the star activity (FWHM, Rhk, bisector span) shows the contribution of the rotation higher harmonics to the observed activity. However, they were of little help to decide on the kind of filtering to be used. The power spectra of these parameters in the period BJD 2,454,847 -- 873 are shown in the fig.~\ref{fig:fig4}. Unfortunately, the features of the spectra of the bisector span and FWHM show only small bumps at the position of the higher-order harmonics, but similar bumps were found when using scrambled data thus showing that the observed one cannot be distinguished from bumps generated by white noise and are thus meaningless. The corresponding data seems to be affected by the low brightness of the star. The only power spectrum showing significant peaks above the minimum level of significance is the power spectrum of the index $\log R_{hk}$ where peaks corresponding to higher-order harmonics are clearly seen even if some offset due to the short time span of the observations used can be noted. A similar analysis using the photometric observations done by Queloz et al.\cite{queloz} (QBM) showed almost no influence of the $4^{th}$  harmonic and led them to neglect it in the construction of the high-pass filter. We will use both filters with and without the fourth harmonic and consider them in the composition leading to the conclusions of this paper on the mass of the planets. In addition we mention that the power spectra clearly show some higher-order harmonics which may affect our results. The consideration of them would require new improvements, different of those presented in this paper.}\\
We shall mention that the use of the rotation's fourth harmonic raises some critical
questions. Indeed its period is one fourth of 23.64d, that is 5.917 d, and one of the main
aliases of the period of 7b is 5.925 d. However, aliases are defined for uniformly spaced
time series with observations separated by multiples of one constant value (e.g. one sidereal
day) and cannot be avoided as far as the separation between observations is kept unaltered.
This is a classically known fact and it has been taken into account in the scheduling of the
spectroscopic observations as soon as the observations showed this coincidence. The
observations done in the period JD 2,454,847 --873 include 52 data obtained in 27
consecutive nights and in 10 of these nights 3 observations were done covering about 4 hours.
This was not sufficient to completely eliminate aliasing problems, but power spectra
extended beyond the nominal Nyquist frequency showed that the perfect mirroring of the
power spectrum has been avoided ({ see fig. 7;} see fig. 10 of QBM). 
Several other tests were done.
Least squares determinations involving simultaneously the photometric period 0.853585 d
and the fourth rotation harmonic lead to correlation values in the range 0.70 -- 0.88
depending on the design of the experiment done and the observations used. Some of these
results are worrisome, but the final test is provided by the coincidence of the convergence
points in chains starting at very different mass values (as the test ones shown in fig.~\ref{fig:fig3}).
One characteristic of least-squares procedures involving highly correlated parameters is that
the results become erratic. This has never been the case here. However, it is clear that
without the multiple-data per night policy adopted in the last periods of observations, it
would be absolutely impossible to separate the rotation's fourth harmonic from the alias of
the period of CoRoT-7b.\\
{In what concerns the higher harmonics, we mention that} the proximity of the period of the $6^{th}$  harmonic (3.94
days) to the period of CoRoT-7c (3.698 days) is a problem of difficult solution. They are
far enough to be separated one from another, but the beat period of the two components
(about 60 days).is much larger than the timespan of the used subset (27 days). When the
algorithm used in this paper is extended to include these harmonics, its convergence becomes
excessively slow. The Fourier analysis of the residuals
obtained with the 4-harmonic high-pass filter and an extended time interval (BJD 2,454,847 -- 884)
indicated that the amplitude of the radial velocities due to CoRoT-7c may be affected in up
to 0.5 m.s$^{-1}$. This possible offset will be taken into account in the final results.
\section{The mass of the planets}
\begin{figure}
\resizebox{\hsize}{!}{\includegraphics{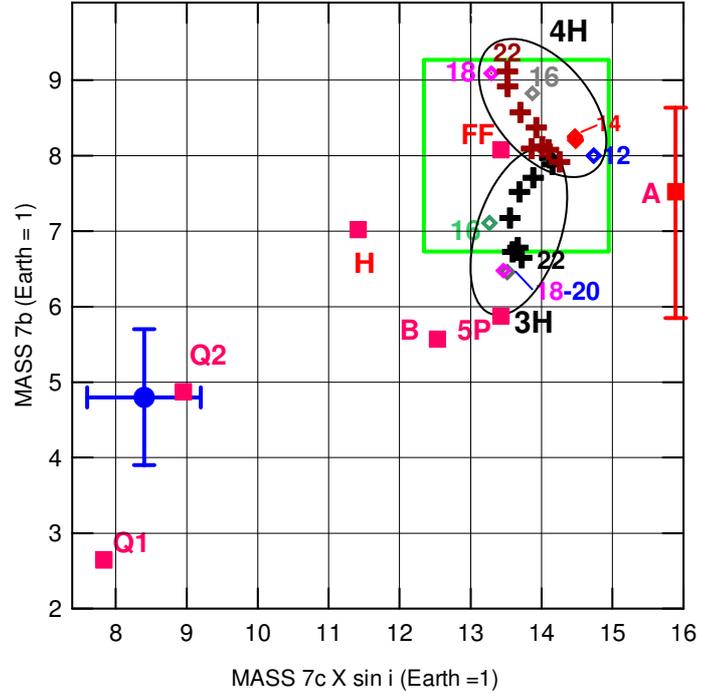}}
\caption{Planet masses resulting from several runs of the iterative high-pass filtering
with 3-, 4-harmonic filters (sets 3H, 4H). The other labels indicate the Coherence
Time in days (see details in the text). 
{The green square represents the interval of confidence of the solution given in Table 2.} 
The large blue cross indicates the adopted solution in QBM and the
red squares Q1 and Q2 the solutions corresponding to the published values of $K_1$,
$K_2$, in QBM. The solutions A (alternative; only CoRoT 7b), B(Boisse), H(Hatzes), 5P (Fourier with 3 harmonics), FF(free Fourier) will be discussed in sections 8 and 9.
}
\label{fig:fig5}
\end{figure}

{ The technique described in the previous sections is certainly the best one we can devise to eliminate from a given series of unevenly spaced observations the contributions of irregular long period terms and to get a remaining part which may be used to determine the parameters of a short-period signal (the planets).}
It is now applied to the observations. However,
for the reasons discussed above, we give up using all observations, but concentrate on the
set of observations done in the period BJD 2,454,847 -- 873. In addition to the aliasing
problem, the consideration of the whole set -- formed by 5 different subsets spanning 106
days plus 3 isolated observations one year before -- is made difficult by the irregular
variation of the stellar activity from the epoch of one set to the next.\\
Another important setting in this determination is that we will concentrate on the masses
and fix the periods in the values previously determined (L\'eger et al. \cite{leger} and QBM). One
of the reasons is that having restricted the interval under study to a set of only 27 days,
there is no possibility of improving a period determination resulting from observations
taken from a time interval 4 times larger. The analysis of the covariance matrix with the 27-
day data shows very high correlation (0.97) between periods and phases for both planets
what means that this short set cannot be used to simultaneously determine periods and
phases.\\
The results of this algorithm depend on the model used in the high-pass filter. Two
main parameters were investigated. One of them is the model used in the interpolation to
determine the activity at a given date: here, we considered both the 3- and the 4-
rotation harmonics models. In a lesser extent, the 6-harmonics models has also been
considered but the beat of the periods of CoRoT-7c and that of the $6^{th}$  rotation harmonic impairs the
procedure convergence. The other parameter is the Coherence Time, which sets the size of
the window. We investigated several of them starting with the 20-day interval as used by
QMB, but considered also some other values in the range 8--22 days. The results 
are shown in fig.~\ref{fig:fig5}. As far as the Coherence Time is kept in a limited
interval, the results do not show large variations. Also, since the
codes themselves depend on some operational parameters that might affect the results, the
procedures of filtering and mass determination were done with two very different codes:
one, a lengthy steepest descent (diamonds) and the other a two-part code using a
genetic algorithm completed with a downhill simplex (crosses). 
{ Labels indicate the corresponding Coherence Times. For the set shown by crosses, only the highest label (22) was shown to avoid excessive overlap 
of symbols and labels; the solutions with Coherence Times between 8 and 12 days cluster around $m_B=8 m_{\rm Earth}$. The others lie between these two limits. 
The green rectangle indicate a joint interval of confidence. 
The individual statistical errors of the self-consistent 
determinations were estimated as 0.5 m.s$^{-1}$ that is, 0.7 and 1.1 Earth masses, for 7b and 7c respectively.}
\\
The sets of masses obtained for both planets show significant dependence on the used high-pass filter. The
mass of CoRoT-7b shows some variations, following the $4^{th}$ rotation harmonic is included or not in
the model. When the 4$^{th}$ harmonic is included, we obtain for CoRoT-7b a mass above 8 
Earth masses, while the results with only 3 harmonics is smaller than 8 Earth masses,
On its turn, the models with 3 and 4 harmonics indicate, for CoRoT-7c, a mass around 14
Earth masses, but the results becomes less well determined when higher harmonics are
included, in a way leading to believe that the actual mass of CoRoT-7c is smaller than the
obtained value. The results of some runs using more observations (the 72 observations done
in the interval BJD 2454845 -- 873), also result for CoRoT-7c a mass smaller than 13
Earth masses. 
{ The results depend also on the period adopted for the planets. Because of its very short period, CoRoT-7b shows a larger sensitivity.} Fortunately, the period of CoRoT-7b is very well known from the CoRoT photometry and this source of error can be discarded.\\
The experiments done have also shown that the results are sensitive to the adopted
weighting rules. All results in this paper were obtained using, at the beginning, the standard
errors published by the observers (L\'eger et al.\cite{leger}), which were propagated following the
classical rules of the least-squares formulas for unequally weighted observations (see
Linnik,  \citep{linnik}) \footnote{The used classical weighting rules are able to take into account the fact that the observations done at the
dates 2454860.75 and 2454864.63 have a quality worse than the others; the weight associated to them was
some 10 times less than the weight given to the more precise observations in the set.}. 
In the sequence, variances were obtained for each subset used in the
filtering and were used to weight the averages on each date giving the estimate of the
activity. At last, let it be reminded that the mass determination from the filtered radial
velocities cannot use the same weights as the given observations. The filtered RVs are
differences between the observed RV and the estimated activity and are, therefore, affected
by the errors in both these quantities. The classical formulas after which the variance ($\sigma^2$)
of the difference is the sum of the variances of the two quantities entering in the subtraction
is used and the new weights are defined as the inverse of the resulting variances.
\section{The star activity}
\begin{figure}
\resizebox{\hsize}{!}{\includegraphics{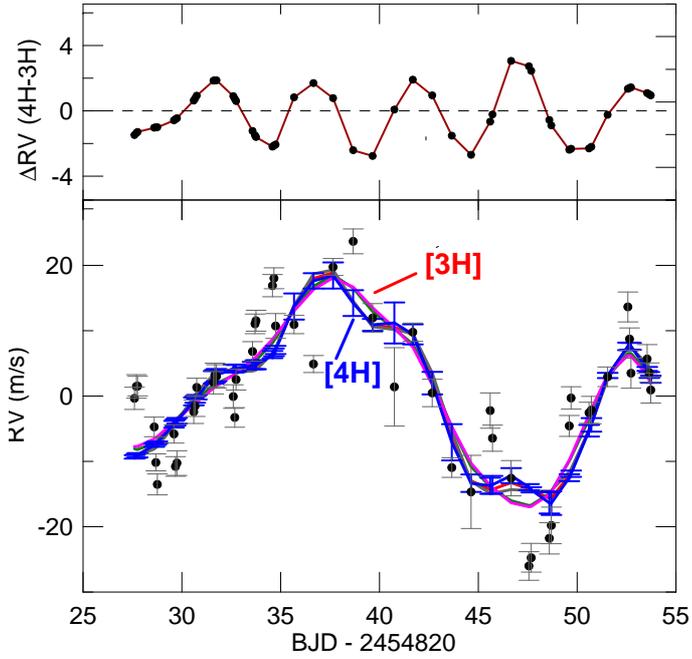}}
\caption{Bottom: Star activity resulting from the high-pass filter. { The labels [4H] and [3H] indicate the results obtained respectively with 4 and 3 harmonics. The color coding of the components of these curves, when visible, is the same as in Fig. \ref{fig:fig7}. The dots indicate the
measured radial velocities with their error bars. The error bars of the activity estimated with the 4 harmonics high-pass filter are also shown. Top: Difference between the results labeled 4H and 3H}}
\label{fig:fig6}
\end{figure}

Figure~\ref{fig:fig6} shows the activity determined using the 3-, 4-harmonic high-pass filters  {(bottom) and the difference between them (top).}. The activities
determined with both filters and various values of the Coherence Time are
shown. For a given model, they do not show visible differences, notwithstanding the fact that these
differences exist and affect the mass determination. The analysis of this figure may be
summarized in a few words: (1) The filtering is very robust with respect to the chosen
Coherence Time and model; (2) The results are more smooth when less harmonics are included in the filter. \\
The activity estimated here may be compared to the one estimated by Pont et al. \cite{pont}
from the analysis of the bisector span measurements. It is remarkable that the features of
the activity given if Fig. 2 of Pont et al. are very similar to those shown in fig.~\ref{fig:fig6} and corresponding to the
4-harmonic filtering; however one may note that the total span of the RV due to the star
activity is, there, about half of that shown in fig.~\ref{fig:fig6}.
\section{Quality of the new high-pass filter}
\begin{figure}
\resizebox{\hsize}{!}{\includegraphics{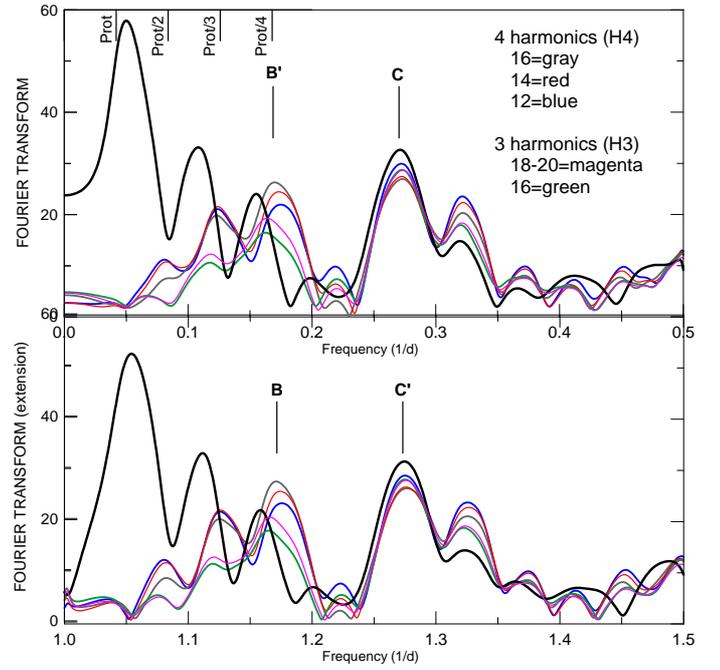}}
\caption{Top: Fourier Transforms (DCDFT) of the observed (black) and filtered data with
the 3-and 4-harmonic high-pass filters in the period BJD 2,454,847 – 873.{
Bottom: Extension of the transforms to beyond the Nyquist frequency to show their behavior below 1 day. $B,C$ are the orbital frequencies of CoRoT 7b and 7c; $B',C'$ are aliases of $B,C$. The rotation period and its first harmonics are indicated on the top axis.}}
\label{fig:fig7}
\end{figure}

As done for the QBM determination, we may compute the Fourier transforms of the
residuals $V_{obs}-V_{activ}$ and compare them to that of the observations. 
{ Some Fourier transforms (DCDFT) are shown in fig.~\ref{fig:fig7} (top) for the two models and several values of the Coherence Time. In order to avoid the complication arising from many almost overlapping curves, we present only those transforms for which the height of the peak near the frequency labeled B' (alias of the orbital frequency of CoRoT-7b) nearly matches the peak of the transform of the observed data.} It is worth stressing that the frequency B' is not separated from the side lobes of the rotation period (which are broad because of the short time span of the used observation set). Because of this superposition, it is difficult to assess the quality of the filtering by inspection of the filtered spectra at this frequency. We may remember that the composition of two frequencies in a spectrum is not just an addition since each of them carries one phase and the effect of the superposition cannot be assessed only by comparing their moduli.
In this case, the superposition is reinforced by the short time span of the considered data. With a longer timespan, peaks would be sharper (as in fig.~\ref{fig:fig2}) and could be separated, but using all available data in the analysis would mean to work with a discontinuous set of observations, which introduces additional (and in some extent unsolvable) difficulties in the estimation of the activity. {One noteworthy effect of the superposition is the apparent enhancement of the peaks at B and B'.}  
\\
The frequency of CoRoT-7c, however, is less affected, at least as far as higher-order
harmonics are not included.\\
{ For the sake of giving an additional information on the aliasing effects, we present in fig.~\ref{fig:fig7} (bottom) an extension of the transforms to an interval of frequencies including the actual orbital frequency of CoRoT-7b (labeled B) and one alias of the orbital frequency of CoRoT-7c (labeled C'). As expected, the transforms in the given intervals are almost identical in both plots, but not perfectly identical because of the large proportion of nights with multiple observations in the selected time interval. \\
The comparison of the Fourier transforms shows that the best quality filtering were obtained with Coherence Times 12-14 days when using the 4-harmonic filter and 18-20 days when using the 3-harmonic filter (the curves for 18 and 20 days are almost identical).}\\
 
\section{The filtered radial velocities}
\begin{figure}
\resizebox{\hsize}{!}{\includegraphics{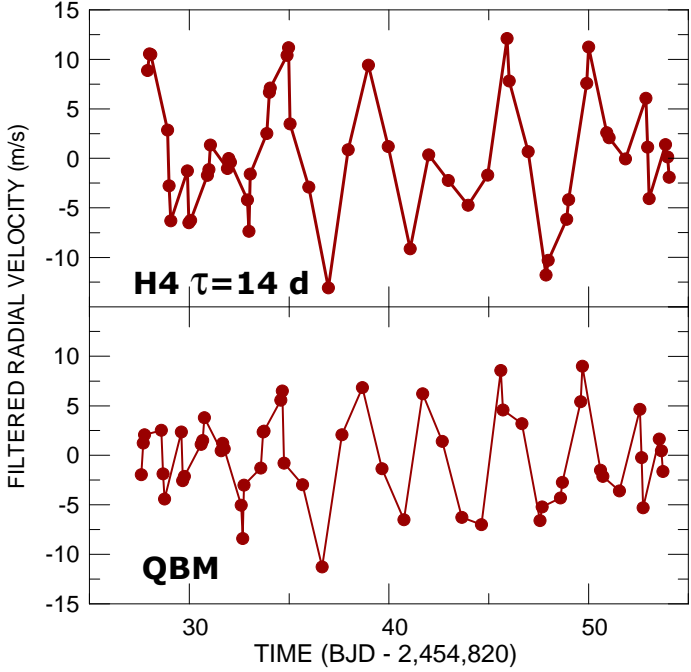}}
\caption{{  Filtered radial velocities obtained with the 4-harmonic high-pass filter and Coherence
Time $\tau = 14 d$ (top) compared to those given in QBM (bottom).}}
\label{fig:fig8}
\end{figure}
Figure~\ref{fig:fig8} (Top) shows the radial velocities obtained subtracting from the measured radial velocities
the activity resulting from the use of the 4-harmonic filter and Coherence Time $\tau = 14 d$.
They may be compared to the ones used in the discovery paper (Fig.~\ref{fig:fig8} Bottom). 
The main differences appear in the beginning of the interval, where the
filtered radial velocities appear much larger in our results than in QBM and in the middle of the interval
where the contrary occurs. It can be easily seen that around the date 42, in our results, a
destructive beat occurs between the RV sine curves corresponding to the two planets, at the
actual observing times.\\
We should stress the fact that the epoch BJD 2,454,820 used in the plots 
and calculations of this paper is not the same epoch used in other papers. It is an arbitrary epoch close to the
actual dates and allows the phases (i.e. the longitudes at the fixed epoch) to depend less strongly on
the periods, thus making the numerical procedures more robust. {In adition, we note that in all steps one additive constant representing the radial velocity of the planetary system is determined together with the other unknowns.  }
\section{A model-independent approach}
\begin{figure}
\resizebox{\hsize}{!}{\includegraphics{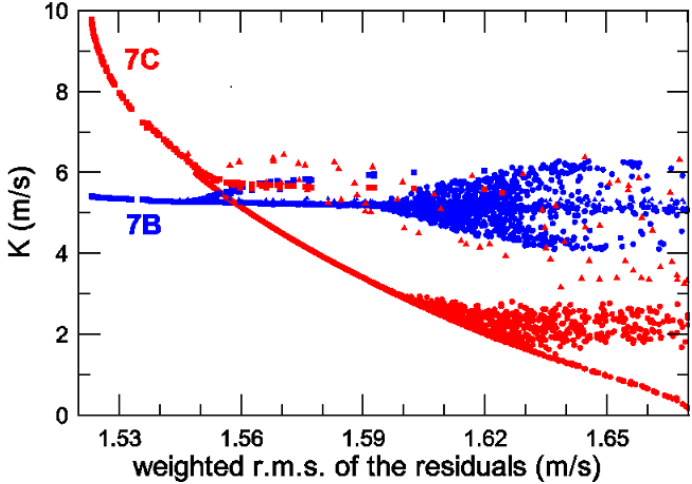}}
\caption{Good fits obtained with a biased Monte-Carlo {under the condition $wrms \leq
1.67$ m.s$^{-1}$}. The blue dots correspond to CoRoT-7b and the red ones to CoRoT-7c.}
\label{fig:fig9}
\end{figure}

 The use of a high-pass filter allowed us to separate parts of the RV measurements due to
high and low frequency components. The results are, however, model-dependent. The
high-pass filtering has given statistically coherent estimates for the mass of CoRoT-7c, but
that result can carry some systematic effects due to some higher-order rotation harmonics.
In what concerns CoRoT-7b, the filtering led to two different solutions. From the purely
statistical point of view, the 4-harmonic filtering should be better than the 3-harmonic
filtering. However because of the aliasing involving the rotation $4^{th}$  harmonic and the
orbital period of CoRoT-7b, it is convenient to confirm the results with some alternative
technique independent of assumptions involving the rotation period and its harmonics. An
alternative approach was suggested by Hatzes et al. \citep{hatzes} (and paper in preparation) in which only data from nights in
which multiple observations were done are used, and assuming that in the 4 hours time
span of one observation night, the activity of the star does not change and may be fixed as
the same for the three observations \footnote{The results from the used RV modeling indicate that 
variations are smaller than the observational errors.
However, in one extreme (and infrequent) case, the estimated activity varied by 0.8 m.s$^{-1}$ between the first and
the last observation done in the night.} . We note that there are 10 nights in the
interval 2,454,847 –- 873, in which three consecutive observations were done. Then, the
problem is to fit those observations with 2 planetary sine waves (the eccentricities may be
assumed equal to zero as discussed in section 10) and a set of 10 independent constants
added to the data in the corresponding dates. The main problem with this approach is the
small number of degrees of freedom in the best-fit problem. We have 30 data and 14
unknowns. The number of degrees of freedom is 16. { In analogy with a chi-square
distribution, we may guess that every solution leading to a w.r.m.s $\leq$0.6 m.s$^{-1}$ higher than the minimum shall be considered as belonging to the standard interval of confidence (see Press et al. \citep{Press}).
In order to improve the determination, we have chosen to add to these observations those
from 16 other nights in which 2 observations were done. We have then 62 data and 30
unknowns. The number of degrees of freedom becomes 32. This is twice of what we
would have if only using the 10 nights with 3 observations each and means a less broad
confidence interval, including only the solutions leading to a w.r.m.s. $\leq$0.16 m.s$^{-1}$ larger than the minimum.}\\
To assess the set of all these 'good-fit' solutions, we may use the same biased Monte-Carlo
technique used to obtain good fits for the planets HD 82943 b,c (Ferraz-Mello et al.  \citep{sylvio1}).
As before, the frequencies were fixed at the values given in QBM (they are well determined) and only 4 unknown
planetary parameters were considered: The two masses and the two phases. The (biased)
random search produced thousands of solutions. { Those with $w.r.m.s. \leq 1.67$ m.s$^{-1}$} are shown
in fig.~\ref{fig:fig9}. In that figure we superpose the results for the half-amplitudes K of the two
planets. { The mass of CoRoT-7b is constrained to the interval $m_B=7.2 \pm 1.4 m_{\rm Earth}$ (i.e.$4.1 \leq K \leq$ 6.2 m.s$^{-1}$) and its
longitude at the epoch BJD 2,454,820.0 is $200 \pm 7$ deg.
The mass of CoRoT-7c, however, is not constrained by this set of
observations. Fits with $w.r.m.s. \leq 1.67$ m.s$^{-1}$ exist for masses in a broad interval: $m_{C} \sin i \leq 22 m_{\rm Earth}$  ($K_C \leq 9.8$ m.s$^{-1}$).}\\
The poor constraining of the solution for CoRoT-7c deserves some discussion. The
alternative technique proposed by Hatzes et al. \citep{hatzes} is good to separate low and high-frequency
contributions. The period of CoRoT-7c is not short enough to be considered as
low. In the more favorable situations, the variations of the RV due to CoRoT-7c may reach
2 m.s$^{-1}$ in 5 hours. It is generally less, and not distinguishable from contributions coming
from low-amplitude long-period terms. We cannot avoid the random process used from
mimicking, in the activity, contributions with the same period as CoRoT-7c and from
combining them with the planet RV. \\
The results are not accurate but they are important because of their independence with
respect to assumptions on the activity behavior.\\
The best constrained fit, defined as the leftmost point in fig.~\ref{fig:fig9} corresponds to the mass $m_B = 7.5$ Earth masses. 
This fit and the interval of confidence corresponding to the maximum span of mthe solutions oin the considered interval is shown by point A in fig.~\ref{fig:fig5}.\\
It is important to stress that each set of masses and
phases of the two planets obtained in this experiment corresponds to one solution whose residuals fit the
observations better than 1.67 m.s$^{-1}$ (w.r.m.s.) and cannot be discarded. The density of the
points in fig.~\ref{fig:fig9} has no meaning (it rather reflects the strategy used to construct the sets).
 We note that only the error bar for $m_B$ is given. It was not possible to get a good determination for $m_c\sin i$ (the error bar would extend more or less over the whole width of the figure).
\section{Fourier approaches}
As a first guess, one could assume that the observations can be represented by the sum of
several periodic functions and use conventional non-windowed Fourier analyses to determine their parameters
(amplitudes, periods, phases). The solutions indicated by B and H in fig.~\ref{fig:fig5} are those
obtained by Boisse et al.  \citep{boisse} and Hatzes et al.  \citep{hatzes} from the Fourier analysis of the
observations. Boisse et al's solution is the best fit of 5 sine curves to the radial velocities, 3
of them having the periods of the rotation and its first harmonics. For that solution,
besides the 52 points used in this paper, the remaining 9 points of the latest observational
subset were also considered. Hatzes et al's solution was obtained using all observations and
a pre-whitening plus filtering procedure based on periodograms; 9 different sine curves
contributed to the construction of the solution.\\
If we proceed in the same way as Boisse et al. \citep{boisse} but considering only the set of 52
observations done used in previous sections, the result is that shown by the symbol 5P in
fig.~\ref{fig:fig5}. In fact, we have done a great deal of Fourier analyses of the measurements, which
remained unpublished only because they assume a periodic behavior in the activity which is
very improbable. Every function in a finite interval may be represented by a Fourier series,
but in the present case, the main period needs to be the rotation period. Such periodic
structure is not seen in the residuals published by QBM (shown in fig.~\ref{fig:fig1} above), in our
own activity curve (shown in fig.~\ref{fig:fig6} above), in the activity as determined by Pont et al.(  \citep{pont}; their fig. 1) 
or in the photometric series produced by CoRoT (L\'eger et al. \citep{leger}). {In all these plots there is some kind of repetition associated with the rotation period, but in some sections the curve appears dominated by the rotation period, in others by the rotation harmonics, and so on. 
None of the partial curves appearing in fig.~\ref{fig:fig1} shows the 23.64 period indicated by photometry. An irregular behavior is also seen in the light curve,  {whose periodogram clearly shows the rotation period and its harmonics, but which is such that} in some sections of it no variations are seen, while in others the rotation period is well marked.}\\
We report here only one of the experiments  {which consisted of several steps:
(a) The periods were initialized in relatively broad intervals bracketing the rotation period and its harmonics, allowing for possible differences due to either the physics of the activity or the beat between the actual periodic signal and the sampling set (actual observation dates).
(b) The approximated more probable elements (including the periods) were determined via a chain of best-fits to the data;
(c) In each step, the elements were determined by the simultaneous best-fit of N trigonometric components.
The process was stopped when the addition of one new term was no longer able to improve significantly the results (F-test) and the scrambled data produced spectra with peaks of the same size as the ones obtained with the real data.
In such case,
the inclusion of more unknowns in the process may lead to high correlation between the unknowns
and to results that may become undistinguishable from artifacts.}\\
The eventual results of the free Fourier approach are shown in fig.~\ref{fig:fig5} (point FF) and
detailed in Table~\ref{table:1}. The error bars were estimated with a biased Monte Carlo sampling of the neighborhood of the solution, which showed very asymmetric distributions in some cases.    \\
\begin{table}
\caption{Fourier decomposition of the data in the time interval
2,454,847 -- 873. The resulting w.r.m.s of the residuals is 1.76 m.s$^{-1}$}             
\label{table:1}      
\centering                        
\begin{tabular}{c c c c}        
\hline\hline                 
Period (d) & $K$(m.s$^{-1}$) & Mass (Earth=1) \\  
    &            &   $\times\sin i$\\  
\hline                        
   21.16 $\pm$ 0.6      & 15.0$^{+0.6}_{-1.4}$  &         \\     
   3.70 $\pm$ 0.04      & 6.0 $\pm$ 0.9         & 13.4$\pm$ 2.0      \\
   0.850 $\pm$ 0.002    & 5.8 $\pm$ 1.2         & 8.1 $\pm$ 1.5     \\
   11.7 $^{+0.3}_{-0.7}$& 4.0 $\pm$ 1.0         &            \\
   5.09 $\pm$ 0.15      & 2.4 $^{+1.4}_{-0.9}$  &             \\ 
\hline                                  
\end{tabular}
\end{table}
 {It is important to stress the fact that the weakest of the five terms is already very uncertain as both the spectrum of the residuals in the previous step and the a posteriori F-test have shown that no actual improvement was obtained by the addition of this term to the solution.}\\ 
One important by-product of this analysis concerns the conjecture of the existence
of one third planet at $P=9.2 d$, which cannot be studied with the other techniques described in
this paper, since they do not allow such a slow periodic variation to be distinguished from
the star activity. Peaks corresponding to periods around 9 days are recurrent in all analyses
done, since the beginning of this investigation. They can be seen for instance in the spectra
of the raw data in fig.~\ref{fig:fig2} and fig.~\ref{fig:fig7}. However, always, when a monochromatic filter is used
and a sinus curve with the amplitude and period of the rotation is subtracted, that peak
disappears. We conjecture that it results from a complex beat between the rotation period,
the sampling dates and the planets themselves.\\
\section{The eccentricities}
\begin{figure}
\resizebox{\hsize}{!}{\includegraphics{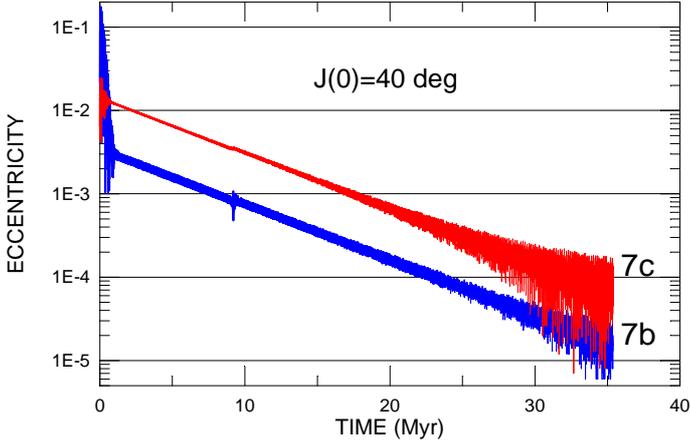}}
\caption{Orbit evolution of two planets with masses 8 and 15 Earth masses in a
CoRoT-7-like system with mutually inclined orbits. The damping of the
eccentricities is due to the tides on the inner planet only.}
\label{fig:fig10}
\end{figure}

In the mass/orbit determination presented in this paper, the eccentricities were taken equal
to zero. Indeed, an analysis of the dynamical problem shows that the tidal dissipation in the
planet CoRoT-7b and its gravitational interaction with CoRoT-7c damp both eccentricities.
Simulations were done where the forces due to the tide raised in the planet (cf Mignard  \citep{mignard}) were
added to the gravitational ones. In all of them, the osculating eccentricities of
the two planets stabilize in values of the order of resp. $10^{-5}$ and $10^{-4}$ in a few tens of Myr,
whichever initial eccentricities and inclinations are considered. In the solution shown in
fig.~\ref{fig:fig10}, the planets are initially on circular orbits with semi-major axes 0.0175 and 0.0456
AU (like CoRoT-7c) in two planes with a mutual inclination of 40 degrees. The chosen
initial eccentricities do not affect the solution because, immediately after the beginning of
the simulation, the gravitational interaction of the 2 planets forces the eccentricities to be
larger than 0.1 and 0.01 respectively (In the coplanar simulations they jump to the
equilibrium values fixed by the mutual perturbations; cf. Mardling  \citep{mardling}; Rodr\'iguez  \citep{adrian}; Rodr\'iguez et al. \citep{adrian1}); Thereafter, the tidal dissipation in the inner planet starts dissipating
the energy of the system, making the orbit of CoRoT-7b slowly spiral down towards the
planet and become circular while the semi-major axis of CoRoT-7c remains almost
unchanged. In addition, the exchange of angular momentum between the two planets also
damps the eccentricity of CoRoT-7c and drives the planets to an equilibrium configuration
(see Mardling  \citep{adrian1}). It is worth recalling that after the circularization of the orbit of
CoRoT-7b, the tidal friction in this planet almost ceases (see Ferraz-Mello et al.  \citep{sylvio2}) and
no longer continues to significantly affect the evolution of the system. Let be added that in
some coplanar runs starting with eccentric orbits, the final eccentricities are yet smaller
than the ones shown in fig.~\ref{fig:fig10}.
\begin{figure}
\resizebox{\hsize}{!}{\includegraphics{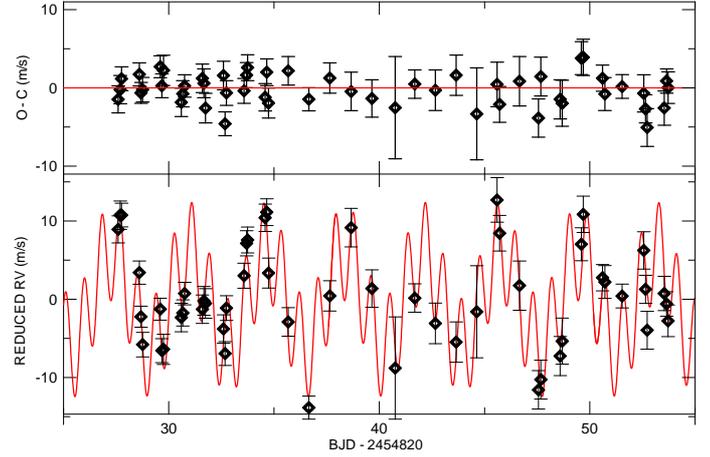}}
\caption{Bottom: Filtered radial velocities with the 4-harmonic high-pass filter and 14-day
Coherence Time (dots) and RV curve corresponding to the planets solution obtained with
the same parameters. Top: Differences between the filtered RV and the RV curve.}
\label{fig:fig11}
\end{figure}
\section{Discussion}
Three independent approaches were used to assess the masses of the exoplanets CoRoT-7b and
CoRoT-7c notwithstanding the difficulties created by the activity of the star which
contaminates the measured velocities not as a "jitter", but as a dominant signal some 2 -- 3
times more intense than the planetary contributions. Before comparing the results, it is
worth stressing the fact that the residuals of the 3 analyzed models have weighted r.m.s. less than 1.9 m.s$^{-1}$,
what may be considered good given the irregular activity of the star. However, we have to
prevent against using the minimum w.r.m.s. of the models to mutually compare them,
because, by definition, these quantities are minima of the residuals with respect to a
given model and therefore, may be considered as model-dependent. They shall be used only
to make comparisons within a given model.\\
The introduction of a self-consistent algorithm improved the filtering used in
QBM and resulted in masses considerably larger than the ones previously obtained. One
fact influencing the results is the adoption of a 4-harmonic filter in addition to the 3-
harmonic filter. The analysis of CoRoT's photometric series did not reveal the need of
using the fourth harmonic. However, the analysis of the spectrographic parameters
associated with the activity of the star (mainly $\log R_{hk}$) showed the importance of
high-orders harmonics even beyond the fourth. In addition, the fact that the $4^{th}$ harmonic is
very close to an alias of the period of CoRoT-7b, makes its introduction necessary. \\
{ The coincidence between the period of the fourth rotation harmonic (5.91 d) and an alias of the orbital period of CoRoT-7b (5.925 d) is a problem of major concern in this study. We have to stress first that in series of astronomical observations done at the same hour angle it is impossible to solve any aliasing problem. In such case, a linear relationship appears between the two components. We may force at will the value of one component and compensate with the value of the other. There is no mathematical tool able to solve this problem. In this case, given the slight difference between the two frequencies (5.91 and 5.925) and the complexity of the signal, the only way to get one solution is to break the uniformity of the time intervals in the sample by means of observations done in very different hour angles. Unfortunately, with observation from only one observatory this is not easy. Differences of a few hours with respect to the mean are the maximum that could have been obtained in this case. The set of observations selected for this study has 52 observations done in 27 nights, including 10 nights with 3 observations in a 4 hours interval and 5 nights with 2 observations. This is the only hope that we have to solve the beat of the two frequencies before new observations are done. They were indeed separated using a 4-harmonic high-pass filter and the convergence of the self-consistent iteration routes to the same result (see fig.~\ref{fig:fig3}) is an indication that, in the adopted algorithm, they may be considered as independent.} \\
 {One additional comment with respect to the beat comes from the analysis of fig.~\ref{fig:fig6} (top). That figure shows the difference between the estimated activity in the two cases: with  3 and 4 harmonics, respectively. It is an evaluation of the contribution of the fourth harmonic to the estimated activity of the star. The surprise, in this case, is that the apparent period of this contribution is not 5.91 d, but only 5.25 d. Assuming that the rotation period is the same observed by CoRoT, this difference would mean that we are not dealing with one frozen periodic signal, but with a signal that is just nearly periodic: as the evolution of a periodic process whose period and phase are continuously changing. The other possibility is that the period is not the same as the published one. We note that 5.25 is one fourth of the main period found in the free Fourier analysis of the radial velocities, 21.15 days, and that the frequency of the highest peak in  the Fourier spectrum of the data, in fig.~\ref{fig:fig7} is near 0.5 d$^F{-1}$ instead of 0.042 d$^{-1}$, inverse of 23.64 d.\footnote{ {The variation from 23.64 to 21,1 days in the activity period is consistent with differential rotation differences expected in a solar-type star
(see Thomas and Weiss,  \citep{spots}). However, the time span of the observations here considered is not large enough to allow us to give full credit to this result and the question should be reconsidered when new data become available.}} Whatever the reason responsible for this difference, it certainly contributed to the fact  that the 4-harmonic filter succeeded to get one solution with the two components separated, while classical least-squares solution and a covariance analysis using the filtered RV obtained with the published rotation was unable to separate the orbital period of CoRoT-7b from the alias of $4^{th}$ harmonic. As a check, we have done some runs of our codes using this lower period. The results are  masses near the values obtained using the actual values of the rotation period. So they confirm those results and show that the filtering algorithm is robust with respect to variations in the rotation period used in the filter, at least as far as it is close to the adopted ones.   }\\
\begin{figure}
\resizebox{\hsize}{!}{\includegraphics{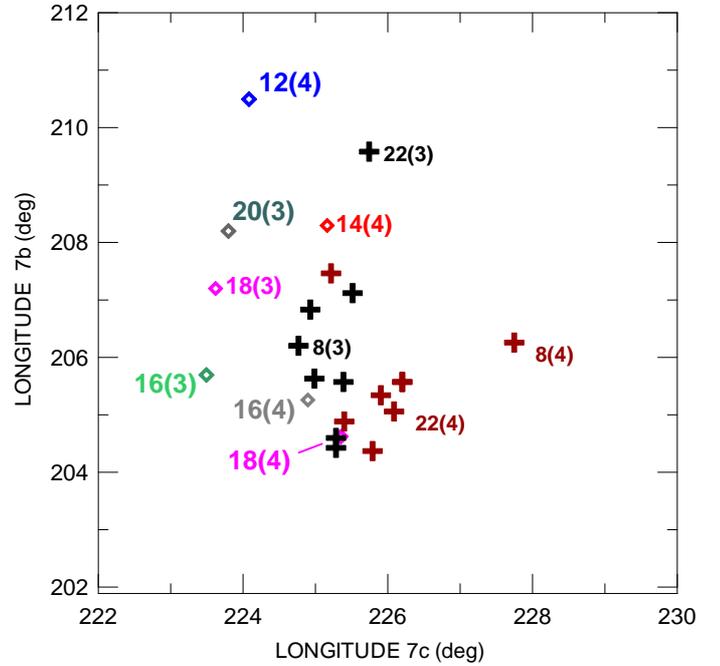}}
\caption{ {Longitudes at the epoch BJD=2,454,820 resulting from several runs of the iterative filtering. The labels indicate the number of harmonics in the high-pass filter and the Coherence Time in days. For the results obtained using the genetic algorithm + simplex code (crosses), only some  results are labeled to avoid overlaps.}}
\label{fig:fig12}
\end{figure}
The quality of the results may be assessed from the residuals (O-C) and from the fitting of
the observations to the model shown in fig.~\ref{fig:fig11}. The obtained (O-C) correspond to a
w.r.m.s. 1.9 m.s$^{-1}$. Two comments may be added here. (1) The error bars in fig.~\ref{fig:fig11} are
larger than those of the RV measurements. They result from the addition of the variances of
the RV measurements to that of the activity given by the used model. The large errors in
some dates in the middle of the interval comes from the fact that, in those dates,
only one observation was done per night, leading to larger statistical errors in the
estimation of the activity in these dates; (2) These results absorb naturally RV jumps as
those that led Pont et al (2010) to state that HARPS systematic errors may be huge in
some cases. One of the 20 m.s$^{-1}$ jumps reported by them, which occurred between the dates
BJD 2454868/69 ($BJD –- 2454820 = 48/49$), is in the studied interval. However,
in our results, it is mainly the consequence of an important increase in the star RV due to the
reaction to the combined motion of the 2 planets at those dates. When the increase due to the activity
(about 3 m.s$^{-1}$) is added to it, one may verify that the part of the RV increase non explained is
still large, but just 1/3 of the increase seen in the RV measurements.\\
An alternative approach suggested by Hatzes et al.  \citep{hatzes}, which has the advantage of
making no explicit hypotheses on the activity (except its slow variation) has also been
improved and combined with a biased Monte Carlo approach. The resulting mass of
CoRoT-7b lies close to results obtained using the 3-harmonic filter. The
resulting mass of CoRoT-7c is ill determined because the period of
CoRoT-7c is not small enough to allow it to be fully separated from the star activity.\\
\section{Conclusions}
The final results were obtained by combining the results of the {3- and 4-harmonic filtering and of the multiple-observations-per-night alternative approach. They are given in Table~\ref{table:2}.
The results from filtering were considered when the two codes used were convergent and lead to nearby values. Otherwise they were not considered in the final estimation of the results. In particular we mention that the largest and least Coherence Times (22 and 8 respectively) were not considered. The other criterion was the quality of the filtering as indicated by the comparison of the Fourier Transforms before and after the filtering (fig.~\ref{fig:fig7}) }
 \\
\begin{table}
\caption{Masses and elements  (Epoch BJD 2,454,820.0) }            
\label{table:2}      
\centering                      
\begin{tabular}{c c c c l}        
\hline\hline                 
          & CoRot-7b & CoRot-7c \\    
\hline               
   K  (m.s$^{-1}$)    & $ 5.7 \pm 0.8$ &  $6.0 \pm 0.6$ \\        
   Mass  (Earth Mass)$\quad (\dag)$& $8 \pm 1.2$ & $13.6 \pm 1.4$         \\ 
   Longitude at Epoch (deg)  & $207 \pm 4$  & $225 \pm 3$      \\
   Period (d) (fixed)  &   0.853585   &   3.698    \\
   Period (d) (estimated) & 0.85354 (see text) & \\
   T0 (BJD) (estimated)	&   2,454,847.893 $\pm$ 0.009  & \\
   Eccentricity (fixed)  &   0   &   0  \\
\hline         
\multicolumn{3}{l}{$\dag$ The mass of CoRoT 7c is minimal ($\times\sin i$)                            }
\end{tabular}
\end{table}
We remind that, because of the small timespan of the observations used, we renounced to determine the
periods and used those given in the discovery papers (L\'eger et al. \citep{leger} and QBM).\\
With the period determined from the transits and the date of the first transit of CoRoT-7b
observed by CoRoT, we obtain 196.4 degrees for the expected longitude at the epoch BJD 2454820.0. The
small offset with respect to the result shown in table\ref{table:2} corresponds to a slight correction in the period, which would then be $0.85354 \pm 0.00002$ days.\\ {
We also stress the fact that the errors given in table~\ref{table:2} intend to define the interval in which the two masses are expected to be. They are not the result of some popular statistical formulas. Those formulas generally are valid in conditions much more restrictive than the actual ones and often the results are much smaller than the actual errors.}
An additional difficulty in the results for CoRoT-7c comes from the fact that it is difficult
to disentangle the mass of this planet from the rotation higher harmonics (mainly the $6^{th}$). In
the case of the beats between the period of CoRoT-7b and the $4^{th}$ harmonic, it was possible
to disentangle them by making multiple observations in many of the nights of the latest
observation periods. In the case of CoRoT-7c, the beats are not related to aliasing, and the
problems they raise cannot be solved by the strategy of making multiple observations per
night nor by mathematical tools allowing low and high-frequency components to be
separated. To solve this kind of entanglement we should have observations spanning over a
larger time interval. We have done some calculations using all observations available in the
interval BJD 2554825 -- 884, increasing the time span from 27 to 60 nights. This means to
include discontinuities in the observations set, which make the analysis more difficult and,
in some sense, of less confidence. In these runs the mass obtained for CoRoT-7c is below
13 Earth masses. Adding this to the evidences from the Fourier analysis of the residuals
used in the runs labeled as 4H, that the $6^{th}$ rotation harmonic may contribute with 0.5 m.s$^{-1}$ to
$K_2$, we may guess that values between 13 and 14 Earth masses are the more probable ones.
One may wish that the improvement of techniques as the one recently proposed by Pont et
al. \citep{pont} applied to simultaneous photometric and spectrographic observation done over
long time spans be able to give a final answer to this question in future. However, with the
existing observations, we are confident that the above estimates are the best ones we can
obtain and that all consistent estimations fall in the ranges given above or, at least, in its
immediate neighborhood.\\
 {The longitudes at BJD=2,454,820 were obtained in the same way as the masses, comparing the solutions from various models (see fig.~\ref{fig:fig12}). 
The individual statistical erorrs are 5 and 12 degrees for planets CoRoT 7b and 7c respectively. The other two approaches did not contribute to the given results because of too large errors.}\\
One last point to note is that the result found here for CoRoT 7b means that 
the planet has a bulk density $11\pm 3.5$  g/cm$^3$. This is much more than the 6.6 g/cm$^3$ resulting from the mass given in QBM and mean not only that CoRoT 7b is rocky, but also that the contribution due to the iron  core must be higher (between 50 and 65 percent in mass; 
see Fortney et al.  \citep{fortney}; Seager et al.  \citep{seager}) and that the density at its center may be close to $25 g/cm^3$ 
(see Seager et al.  \citep{seager}). These values are higher than those corresponding to other known  rocky planets, but all consistent 
determinations of the mass of CoRoT-7b lead to bulk densities of at least $9 g/cm^3$ showing that the high-density of CoRoT-7b 
is a constraint to be taken into account in the modeling of the planet.\\ 
{ The comparison of the various methods used in this paper allows us to say that the high-pass filter used by QBM (Queloz et al. \citep{queloz}), embedded in  a self-consistent algorithm, is the best one we can devise to disentangle long- and short-period terms in a given series of unevenly spaced observations. Its superiority over Fourier analyses with fixed frequencies comes from the fact that the used running window allows the method to treat a signal which is not periodic or has a period different of the period used in the filter. The procedure proposed by Hatzes et al. \citep{hatzes} of using only nights with multiple observations and including an additional $V_0$ for each date is an important complement. It suffers, however, from a shortcoming due to the great number of unknowns that it involves. As a consequence the number of degrees of freedom is small and the resulting confidence intervals are too large. In the current case, it just allowed us to determine the mass of CoRoT 7b. This was great because this was the ultimate goal of our work, but the impossibility of getting reasonable confidence intervals for the mass of CoRoT 7c shall be mentioned. The planning of new observations should take this into account and have as many multiple night observations as possible.} 
\begin{acknowledgements}
SFM and CB acknowledge the fellowships of the Isaac Newton Institute for Mathematical
Sciences, University of Cambridge (UK) where they have developed the tools used to
investigate this problem. MTS acknowledge FAPESP. The continuous
support of CNPq (grants 302783/2007-5 and 485447/2007-0) and the CAPES-SECYT
Brazil-Argentina joint science program (grant 131-07) are also acknowledged. In the course
of this investigation, several members of the CoRoT Exoplanets Science Team have
supplied us with relevant information on the RV data and their analysis. We want
particularly thank A. Hatzes and D. Queloz for their help.{
We are pleased to acknowledge the excellent reviewer report, raising many questions and contributing enormously to the improvement of the paper.}
\end{acknowledgements}

\end{document}